\documentclass[prl,aps,twocolumn]{revtex4}

\usepackage{xspace,amsmath,amsfonts,amsthm,amssymb,amsbsy}
\usepackage{graphicx,subfigure}
\usepackage{amssymb}
\usepackage{color}
\usepackage{psfrag}
\usepackage{subeqnarray}

\definecolor{pink}{rgb}{1,0.078,0.57}

\definecolor{green}{rgb}{0,0.7,0.9}

\newcommand{\ket}[1] {\left\vert #1 \right\rangle}

\newcommand{\dg}{^{\dagger}}

\graphicspath{{../images/}}

\begin{document}

\title{Photocreation of a dark electron-hole pair in a quantum dot}

\author{Shiue-Yuan Shiau}
\affiliation{Physics Division, National Center for Theoretical Sciences, Hsinchu 30013, Taiwan}
\author{Benoit Eble}
\affiliation{Sorbonne Universit\'e, CNRS, Institut des NanoSciences de Paris, 75005 Paris, France}
\author{Valia Voliotis}
\affiliation{Sorbonne Universit\'e, CNRS, Institut des NanoSciences de Paris, 75005 Paris, France}
\author{Monique Combescot}
\affiliation{Sorbonne Universit\'e, CNRS, Institut des NanoSciences de Paris, 75005 Paris, France}
\date{\today}

\begin{abstract}

Photon absorption in a semiconductor produces \textit{bright} excitons that recombine very fast into  photons. We here show that in a quantum dot set close to a \textit{p}-doped reservoir, this absorption can  produce a \textit{dark} duo, \textit{i.e.}, an electron-hole pair that does not emit light. This unexpected effect relies on the fact that the  wave function for a hole leaks out of a finite-barrier dot less than for electron. This difference can render the positively charged trio unstable in the dot by tuning the applied bias voltage in a field-effect device. The unstable trio that would result from photon absorption in a positively charged dot,  has to eject one of its two holes. The remaining duo  can be made dark with a probability close to 100\%  after a few pumping cycles with linearly polarized photons, in this way engineering long-lived initial states for quantum  information processing.

\end{abstract}
\maketitle


The demand for memory storage is fueling the search for long-lived qubits. 
Due to tremendous progresses in nano-fabrication, semiconductor quantum dots (QD) are highly promising stationary qubits based on solid state systems, with applications in quantum communication, quantum sensing and quantum computing nano-devices \cite{livre-Michler}. When trapped in a QD, a carrier spin or a photocreated electron-hole (\textit{eh}) pair implements qubit that can be coherently initialized, manipulated and read out using short laser pulses \cite{Atature,Gao,Warburton-revue}. In the quest for robust long-lived qubits, two strong candidates have emerged: (1) the hole spin which has a coherence time as long as a few hundred  microseconds, due to its weak hyperfine interaction with nuclear spins \cite{Eble,Gerardot,Prechtel,Fras}; (2) the dark \textit{eh} pair \cite{Schwartz2}  because being made of same-spin carriers, it cannot recombine. A dark pair, which should be better called dark ``duo'' instead of dark ``exciton'' for reasons developed in \cite{footnote1}, can stay in a dot for over  $\mu$s, until it turns bright due to spin relaxation \cite{McFarlane,Gershoni} or valence band mixing \cite{Smolenski,Zielinsky,Germanis}. By contrast, a bright duo only lasts  a few hundred picoseconds before recombination \cite{Minnielle}. While the hole spin in a QD provides a qubit that can be deterministically controlled with high fidelity by using charge-controlled devices \cite{Gerardot,Warburton},  the only  scheme based on  dark duos  proposed up to now, relies on the radiative cascade of a metastable biexciton \cite{Gershoni,Schwartz}.

\begin{figure}
\begin{center}
\includegraphics[trim=5.5cm 4.7cm 6cm 6cm,clip,width=2.8in] {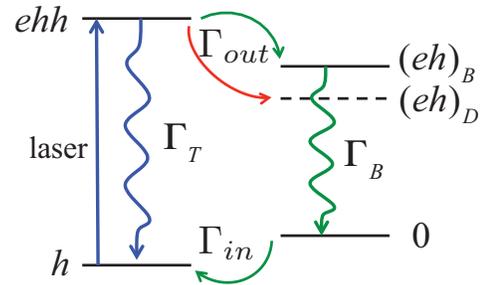}
\caption{Creation of a dark duo (red). The pumping laser is tuned on the $\textit{h}\rightarrow \textit{ehh}$ resonance. The hole tunneling rates in and out  the dot are $\Gamma_{in}\sim \Gamma_{out}$, while $\Gamma_{T}\sim\Gamma_{B}$ are the spontaneous recombination rates of the trio and  bright duos.} 
\label{fig1}
\end{center}
\end{figure}

\textbf{In this Letter}, we propose a protocol that, in a $p$-doped QD structure, deterministically produces a dark duo through a three-step process, the \textit{ehh} trio serving as unstable excited state (see Fig.~\ref{fig1}): (1) an empty dot is charged with a hole  from the nearby $p$-doped reservoir; (2)  an \textit{eh} pair is photocreated in this positively charged dot, which would lead to an \textit{ehh} trio; (3) as the \textit{ehh} trio is unstable, one of its two holes tunnels out. The duo that remains in the dot can be  dark or bright. If it is dark, the dot  becomes transparent to any new incoming photon due to Pauli blocking \cite{transparent}  and stays with its dark duo; if bright, the duo recombines and the above cycle can continue until the dot ends with a dark duo. 

To realize such a cycle, the simplest idea is a device based on a \textit{p-i-n} type diode, with a layer of dots located at  a few ten nanometers from the surface of a \textit{p}-doped reservoir \cite{Gerardot,Brunner,charge-controlled-devices}, in order for the hole tunneling rate to be  comparable to the spontaneous recombination rate of the \textit{eh} pair in a trio, of the order of ns$^{-1}$. 

\begin{figure*}[t]
\begin{center}
\includegraphics[width=2\columnwidth] {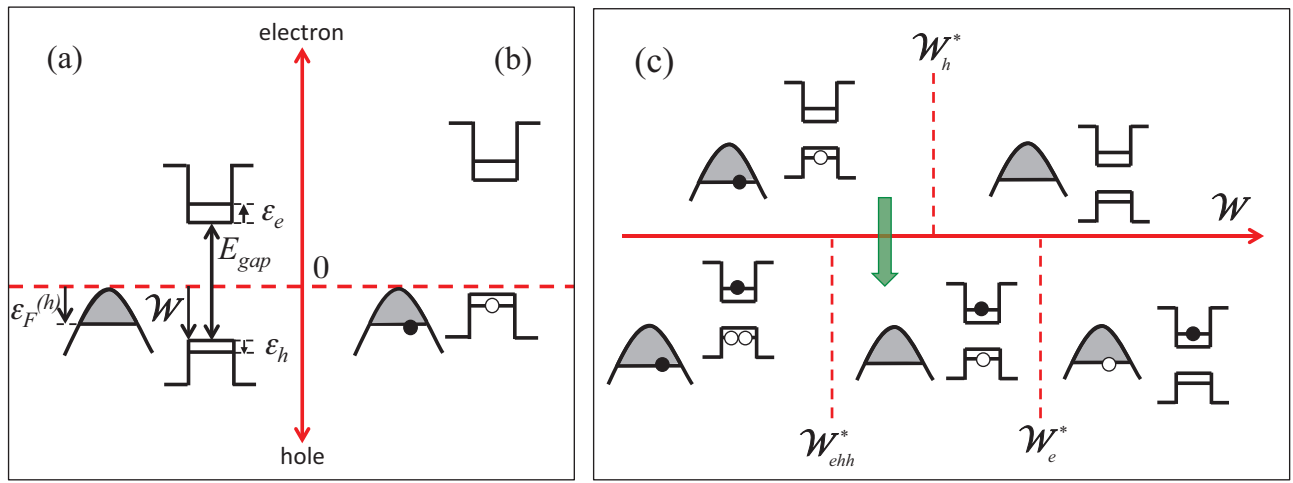}
\end{center}
\vspace{-7cm}
\caption{ \small Without photocreated \textit{eh} pair, the QD is empty (a) when the hole energy level is high, while it contains a hole (b) when the hole level is low.  Electrons  and holes are represented by black and white dots; their energy axes  go in opposite directions. The hole electrostatic energy $\mathcal{W}$  can be changed through a bias voltage between the QD and the \textit{p}-doped reservoir.  (c) Occupancies of the QD and the $p$-doped reservoir as a function of $\mathcal{W}$ without (upper panel) and with (lower panel) a photocreated \textit{eh} pair.  When $\mathcal{W}^*_{ehh}<\mathcal{W}<\mathcal{W}^*_h$, the QD contains a hole before photon absorption but one \textit{eh} pair only after: absorbing a photon (green arrow) then goes along with expelling a hole from the dot. }
\label{fig2}
\end{figure*}

Our proposal relies on the idea that the  dot having a neutral $eh$ pair is positively charged because for finite barrier heights, the  wave function for hole leaks out of the dot less than for electron \cite{SM}. So,  the energy cost for a hole to be trapped in a dot is less when the dot is empty than when it contains an $eh$ duo. Through an appropriate bias voltage, it is possible to make one hole stable in the dot but not an  $ehh$ trio: one hole has to leave when an $eh$ pair is photocreated in a  positively charged dot. Depending on the ejected hole spin, the remaining duo can be bright or dark.

The caveat is that the spin of the tunneling hole is uncontrollable. To overcome this issue, we can repeat the cycle. This demands (i) short pump pulses to avoid  stimulated trio recombination, (ii) the time between pulses  synced to the cycle time, (iii) a fast cycle time   to reach a high dark-duo probability during the dark duo lifetime. Another problem is to never end in a state transparent to the  laser pulse. This can be done by using linearly polarized photons.

Let us now delve into the above physics.

\textbf{\textit{On the \textit{p}-doped structure.}}---
 The cycle we propose requires a reservoir of holes because  holes,  far heavier than electrons,  leak less out of a finite-barrier dot. This stabilizes the \textit{eeh} trio, but destabilizes  the \textit{ehh}  trio \cite{SM}.  By choosing  the  applied bias voltage such that the photocreated \textit{ehh}  trio becomes unstable, one of the two holes has to tunnel out, leaving the dot with a neutral \textit{eh} duo.

Let $\varepsilon_e$ and $\varepsilon_h$ be the energies of an electron and a hole in the QD (see Fig.~\ref{fig2}a). Due to Coulomb contribution, its induced energy denoted as $\gamma$, we can write the energy of an \textit{eh} duo as $\varepsilon_{eh}=\varepsilon_e+\varepsilon_h+\gamma_{eh}$, with $\gamma_{eh}$ negative, and the energy of a \textit{ehh} trio as  $\varepsilon_{ehh}=\varepsilon_e+2\varepsilon_h+\gamma_{ehh}$.  For small dots,  carrier correlations are weak; so,  $\gamma_{ehh}\simeq2 \gamma_{eh}+\gamma_{hh}$. Since in a QD, the  wave function for hole extends less than for electron,  two holes repel each other more than  one hole attracts an electron; so,  $|\gamma_{eh} |<\gamma_{hh}$, which leads to
\begin{equation}
0<\gamma_{ehh}-\gamma_{eh}\,.\label{eq:eeheh<0}
\end{equation}
The sign of this difference  makes one hole   stable in the QD, whereas this is not necessarily so for an \textit{ehh} trio \cite{SM}. In  a few nanometers thick InAs/GaAs quantum dot, experiments \cite{Finley,binding-energies-trions,Regelman} give this difference as a few meV.

\begin{figure*}[t]
\begin{center}
\includegraphics[width=2\columnwidth] {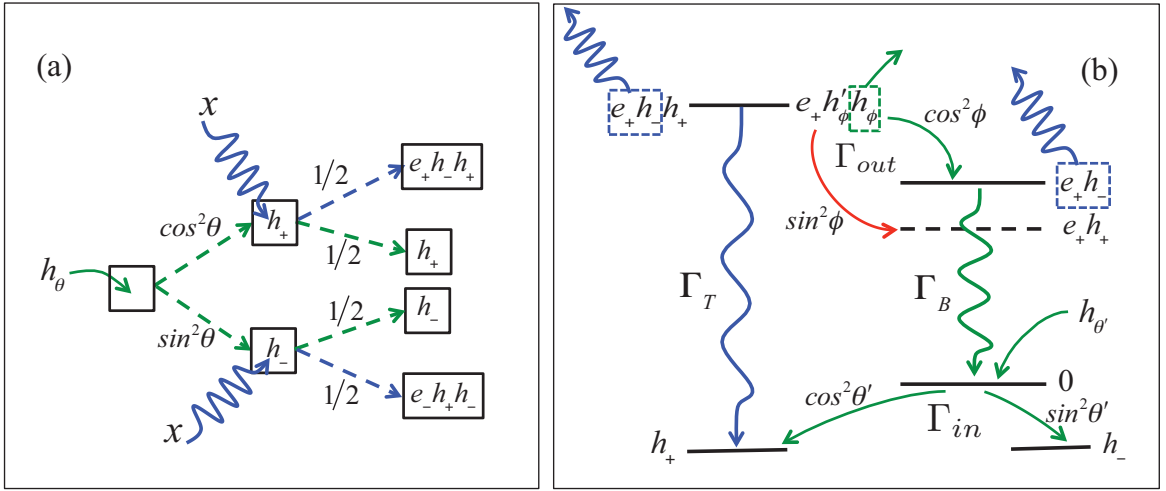}
\end{center}
\vspace{-6.5cm}\caption{\small (a) When a $h_\theta$ hole tunnels  to  the empty dot, the dot hosts a $h_{+}$ or $h_{-}$ hole with a probability $ \cos^2 \theta$ or $\sin^2 \theta$. After irradiated with a short $x$-photon pulse, the dot either stays unchanged  or welcomes a photocreated pair, with an equal probability. This photocreated pair is $e_+ h_-$ or $e_- h_+$ depending on if the dot contains a $h_{+}$ or $h_{-}$ hole. (b) After the end of the  pump pulse, the $e_+h_-h_+$ trio either suffers a spontaneous $eh$ recombination  with a $\Gamma_T$ rate,  or one of its two holes  tunnels out of the dot with a $\Gamma_ {out}$ rate. When the $h_\phi$ hole tunnels out, the dot has a probability $\sin^2\phi$ to stay with a dark duo $e_+h_+$, and $\cos^2\phi$ to stay with a bright duo $e_+h_-$  that recombines with a $\Gamma_B$ rate. The empty dot  then welcomes a $h_{\theta'}$ hole and a new cycle can begin when another pulse arrives. This new cycle generates an $e_+h_-h_+$ or $e_-h_+h_-$ trio depending on if the dot contains a $h_+$ or $h_-$ hole.}
\label{fig3}
\end{figure*}

\textbf{\textit{On the bias voltage.}}---We consider a QD close to a $p$-doped reservoir, the  hole tunneling rate being possibly increased by acting on the bias voltage.

$\bullet$ Without  photocreated \textit{eh} pair, the QD is either empty or contains a hole, depending on the hole electrostatic energy $\mathcal{W}$ induced by the bias  voltage. For $\mathcal{W}$ large, the QD is empty (Fig.~\ref{fig2}a), while below a $\mathcal{W}_{h}^*$ threshold, a hole tunnels to the dot --- which amounts to adding an electron to the reservoir (Fig.~\ref{fig2}b).


 $\bullet$ When an \textit{eh} pair is photocreated, the dot can  have an \textit{ehh} trio, an \textit{eh} duo or just an electron (lower panel of Fig.~\ref{fig2}c). For $\mathcal{W}$  larger than $\mathcal{W}_{e}^*$, the photocreated hole tunnels to the reservoir, and the QD stays with  the electron.  For  $\mathcal{W}$  smaller than $\mathcal{W}_{ehh}^*$, a hole tunnels to the QD to form a stable \textit{ehh} trio.  In between, $\mathcal{W}_{ehh}^*<\mathcal{W}<\mathcal{W}_{e}^*$, the QD contains  one \textit{eh} duo, the hole sea being unchanged.

 As shown in \cite{SM}, these  thresholds are ordered as 
      \begin{equation}
    \mathcal{W}_{ehh}^* < \mathcal{W}_{h}^* < \mathcal{W}_{e}^*\,,
      \end{equation} 
      which follows from the inequality (\ref{eq:eeheh<0}).
      
In the following, we will restrict to  $ \mathcal{W}_{ehh}^* < \mathcal{W}< \mathcal{W}_{h}^*$, that is, a bias voltage for which  the QD hosts a hole, but once an \textit{eh} pair  is photocreated, one of the two holes has to tunnel out.  The $\mathcal{W}_{ehh}^*-\mathcal{W}_{h}^*$ range, equal to $\gamma_{ehh}-\gamma_{eh}$ (see \cite{SM}), is  large enough to experimentally set the bias voltage within this range.

\textbf{\textit{On the spin of the tunneling hole.}}---Due to spin-orbit interaction and  confinement in a dot with growth axis $z$, the involved holes  are the heavy \cite{footnote2} holes $h_\pm=(\pm1)_z \otimes (\pm1/2)_z$ with spin $(\pm 1/2)_z$ and orbital symmetry $(\pm1)_z=(\mp i x+y)/\sqrt{2}$. The $h_\pm$ holes being degenerate in the dot and the reservoir, the  one that tunnels is a linear combination  of $h_\pm$, its creation operator reading in terms of $ h_{\pm}$  creation operators  $b\dg_{\pm }$ as 
    \begin{equation}\label{bh+that}
   b\dg_{\theta}=\cos 
   \theta \,
   b\dg_+         +  \sin \theta \, b\dg_-\,,
    \end{equation}
    if we forget phase factor.

When $\mathcal{W}<\mathcal{W}_h^*$, the  $h_\theta$ hole coming into the empty dot has an unknown $\theta$. So, the probability for the dot to contain a $h_+$ or $h_-$ hole is 
 $\cos^2\theta$ or $\sin^2\theta$, respectively (Fig.~\ref{fig3}a).


\textbf{\textit{On the photon polarization.}}--- We irradiate the QD having a $h_{\theta}$ hole with  photons propagating along the growth axis $z$.
The coupling between these photons and QD carriers  reads 
 \begin{equation}
     W_{ph-dot}= \Omega_t\sum_{\eta=\pm}a\dg_{- \eta }     b\dg_{\eta}\alpha_ {\eta}+ h.c.\,,
       \end{equation}
       where $a\dg_{\pm }$ creates a conduction electron with spin $(\pm 1/2)_z$ and $\alpha_ {\pm}$ destroys a photon with circular polarization $\sigma_\pm$. The time dependence of the Rabi coupling $\Omega_t$ corresponds to a short $\pi$ pulse in order to obtain an efficient   transfer from $h$ to $ehh$ in the dot.\

 $\bullet$ Let us first consider a pulse made of $\sigma_-$ photons. Such a photon is  coupled to $e_+h_-$ pair; so, 
    
 (i) it does not act on a QD  holding a $h_-$ hole due to Pauli blocking: the QD then stays with  $h_-$. 

(ii) it is absorbed by a QD  holding a $h_+$ hole. However, when $\mathcal{W}_{ehh}^*<\mathcal{W}$, the $e_+h_-h_+$ trio is unstable; so, one of its two holes must tunnel out. The key to go further is to note that the trio state $b\dg_{+} a\dg_{+}     b\dg_{-}\ket{v}$, where $\ket{v}$ is the vacuum, also reads 
         \begin{equation}
  ( \cos \phi\, b\dg_{+}+  \sin \phi \,b\dg_{-}) a\dg_{+}  ( - \sin \phi \,b\dg_{+}+  \cos \phi\, b\dg_{-})\ket{v}\,
   \end{equation} 
whatever $\phi$, as easy to check. So, the hole that tunnels out of the dot can be a $h_{\phi}$ hole. The remaining \textit{eh} pair state  
   \begin{equation}
a\dg_{+}  \big(-\sin \phi\,
   b\dg_{+}         +  \cos \phi \, b\dg_{-}\big) \ket{v}                                
   \end{equation} 
then corresponds to a dark duo $e_+  h_+$ with a probability $\sin^2\phi $ and a bright duo $e_+  h_-$ with a probability $\cos^2\phi$. This bright duo will recombine; a new hole will tunnel to the empty dot  and a new cycle can start again.
 
 $\bullet$ The problem  is  that a dot with a $h_-$ hole is transparent to all incoming $\sigma_-$ photons; so, no dark duo can be produced anymore. To avoid this plight, we can use  linearly polarized photons $\sigma_x$. A dot with a  hole, $h_+$ or $h_-$, then has an equal probability to stay unchanged or to absorb a $\sigma_x$ photon and produce an $e_\pm h_+h_-$ trio.  Therefore, the unique dot state transparent to $\sigma_x$ photons is a dot with a dark duo,  which then is the only possible final state after many cycles.

\textbf{\textit{On the trio evolution.}}---After the end of a pump pulse, the \textit{ehh} trio can lose either one of its two holes by tunneling out  at a $\Gamma _{out}$ rate, {\it or} its $eh$ pair  by spontaneous recombination at the $\Gamma _{T}$ rate (Fig.~\ref{fig1}). The    time  evolution of the $e_+h_-h_+$ trio  follows from $d n_T /dt= - (\Gamma _{out}+\Gamma _{T}) n_T$. After the tunneling-out of a $h_\phi$ hole (Fig.~\ref{fig3}b), the  time evolution of the remaining duo follows from $d n_D /dt= \sin^2\phi  \,\Gamma _{out}\, n_T$ if it is dark, and $d n_B /dt= \cos^2\phi  \,\Gamma _{out}\, n_T-\Gamma _{B}\,n_B$ if it is bright, since the bright pair can  recombine. In the latter case,  a new hole tunnels into the empty dot from the reservoir.

From these rate equations, we can derive the various dot occupancies. As shown in \cite{SM}, they read in terms of   
 \begin{equation}
 R=\frac{\Gamma _{out}}{\Gamma _{out}+\Gamma _T}\,,
   \end{equation} 
which results from the  competition between hole tunneling and spontaneous $eh$ recombination in a trio.


\textbf{\textit{Dark duo probability after \textit{n} cycles.}}---The algebra to derive the dot occupancies is greatly simplified if we  consider that the holes which tunnel in and out the dot are ``average'' holes, that is, $(h_++h_-)/\sqrt{2}$. Results for general ($h_\theta,h_\phi$) holes can be found in \cite{SM}. 

$\bullet$ We then start with a dot that has an equal probability to be occupied by a $h_ \pm$ hole, $F_\pm^{(0)}=1/2$. 

$\bullet$ After the absorption of a $\sigma_x$ photon and the evolution of the resulting trio, the dot can contain one of the two  dark duos, $e_+ h_+$ or $e_- h_-$, with a probability $G_\pm^{(1)}= (R/4) F_\pm^{(0)}$ (see \cite{SM}). Accordingly, the dot occupation by a $h_ \pm$ hole reduces to $F_\pm^{(1)}=F_\pm^{(0)}-G_\pm^{(1)}=(1-R/4)F_\pm^{(0)}$. 

$\bullet$ The second cycle, which starts with a smaller hole occupation, brings an additional dark-duo probability $(R/4)F_\pm^{(1)}$; so, the dark duo occupation becomes $G_\pm^{(2)}= (R/4)(F_\pm^{(0)}+F_\pm^{(1)})$, while the hole occupation reduces further to $F_\pm^{(2)}=(1-R/4)F_\pm^{(1)}$. 

$\bullet$ Iteration  to the $n^{th}$ cycle gives the hole occupation as $F_\pm^{(n)}=F_\pm^{(0)}(1-R/4)^n$ and the dark duo occupation as
  \begin{equation}
G_\pm^{(n)}=F_\pm^{(0)}\frac{R}{4}\sum_{m=0}^{n-1}\left(1{-}\frac{R}{4}\right)^m
   =\frac{1}{2}   \left [1{-}\left(1{-}\frac{R}{4}\right)^n\right].\label{Gpmn}
   \end{equation}
 So, after many cycles, $F_\pm^{(n)}\simeq 0$ and $G_\pm^{(n)}\simeq 1/2$, which gives the probability $G^{(n)}=G_+^{(n)}+G_-^{(n)}$ close to 1 to have a dark duo, either $e_+ h_+$ or $e_- h_-$.

The number of cycles required  to reach the stationary regime decreases with increasing $R$, that is, increasing $\Gamma_{out}$, as possibly done by tuning  $\mathcal{W}$ close to $\mathcal{W}_{ehh}^*$.  For its maximum value $R=1$, the probability $G^{(n)}$ to get a dark duo reaches 95\% after 10 cycles and 99\% after 15 cycles. In addition, to  obtain a high dark-duo probability within a span shorter than the dark duo lifetime, it is necessary to have a short cycle time. For a short pump pulse duration, the cycle time scales as
 \begin{equation}
 \tau_{cycle}\sim\frac{1}{\Gamma_ {out}}+\frac{1}{\Gamma_B}+\frac{1}{\Gamma_{in} }\,.
   \end{equation}

To estimate $\Gamma_ {out}$, we use a WKB approximation \cite{Oulton}. For InAs/GaAs dot having a 1.4 nm size, a hole mass $m_h=0.41m_0$, a 40 kV/cm electric field, and a 50 meV hole ionization energy, we obtain $\Gamma_{out}\simeq$ 1 ns$^{-1}$. By considering $\Gamma_{out}\simeq\Gamma_{in}$ and $\Gamma_T\simeq \Gamma_B$ of the same  order of 1 ns$^{-1}$, we have $R\simeq1/2$. Then, the time necessary to produce a dark duo with a $90\%$ probability is of the order of a tenth of  the bright duo lifetime \cite{SM}.

\textbf{\textit{Reading out the dark duo occupation.}}---We can  track the dark duo occupancy by measuring the number of photons emitted by  bright duos,  since the trio and duo emission lines are shifted in energy. This non-resonant read-out protocol is  easy to implement experimentally. The rate equations for the trio evolution give \cite{SM} the probability for  bright duos to emit a photon after $n$ cycles as 
\begin{equation}
L^{(n)}_{ph}= 1-\left(1-\frac{R}{ 4   }\right)^n\,,
\end{equation}
which just corresponds to the dark duo probability,  $G^{(n)}$ (see Eq.~(\ref{Gpmn})).  The number of cycles $n$ is related to the total duration of the pulsed excitation sequence $\tau_L$ and to the pulse repetition frequency $\omega_L\simeq1/\tau_{cycle}$, through $n=\omega_L\tau_L $. Saturation of the   bright-duo photon emission  constitutes a direct optical signature that the QD contains a dark duo.

%
%

\textbf{\textit{Experimental Implementation.}}---By using charge-tunable devices, it is possible to control the occupation  of confined states at the single carrier level \cite{Gerardot,Brunner,Prechtel} and to observe sharp luminescence lines corresponding to one neutral pair, one charged pair and so on, these lines being shifted due to Coulomb interaction.

Fast  generation of dark duos through the  pumping cycle we propose also requires   $\Gamma_{out}$ to be sizable compared to $\Gamma_{T}$. In realistic \textit{p}-doped diodes, there is a bias range where  emission lines from different charged states can coexist \cite{Gerardot}. So,  the threshold $\mathcal{W}_{ehh}^{*}$  can be easily found experimentally. 
This confirms that the external bias can make the QD-reservoir tunneling rate  comparable to the recombination rate, which  exactly is the desired regime for the cycle implementation.

\textbf{Conclusion.}---The protocol we here propose consists of a set of conceptually simple processes that lead to an unexpected effect: the photocreation of a dark electron-hole pair in a quantum dot.

This prediction physically comes from the difference in leakage of the electron and hole wave functions from a dot having finite barriers. For a  bias voltage between the dot and a \textit{p}-doped reservoir chosen such that a hole is stable in a dot, but not a  positively charged trio, the photocreation of an electron-hole pair goes along with the expulsion of a hole from the dot, which can then end with a dark duo. Dark duos are long-lived storage units that cannot recombine into  photons. We hope that the present work will stimulate more experiments on dot-based quantum memories.


 

\end{document}